# Magnetic Electrides: High-Throughput Material Screening, Intriguing Properties, and Applications

Xiaoming Zhang [1, 2, ‡ *], Weizhen Meng [1, 2, ‡], Ying Liu [1, 2], Xuefang Dai [1, 2], Guodong Liu [1, 2*], and Liangzhi Kou [3*]

[1] State Key Laboratory of Reliability and Intelligence of Electrical Equipment, Hebei University of Technology, Tianjin 300130, China.

[2] School of Materials Science and Engineering, Hebei University of Technology, Tianjin 300130, China.

[3] School of Mechanical, Medical and Process Engineering, Queensland University of Technology, Garden Point Campus, QLD 4001, Brisbane, Australia.

‡ These authors contributed equally: Xiaoming Zhang and Weizhen Meng

* Corresponding authors. Email: zhangxiaoming87@hebut.edu.cn, gdliu1978@126.com, liangzhi.kou@qut.edu.au.

**ABSTRACT:** Electrides are a unique class of electron-rich materials where excess electrons are localized in interstitial lattice sites as anions, leading to a range of unique properties and applications. While hundreds of electrides have been discovered in recent years, magnetic electrides have received limited attention, with few investigations into their fundamental physics and practical applications. In this work, 51 magnetic electrides (12 antiferromagnetic, 13 ferromagnetic, and 26 interstitial-magnetic) were identified using high-throughput computational screening methods and the latest Material Project database. Based on their compositions, these magnetic electrides can be classified as magnetic semiconductors, metals, or half-metals, each with unique topological states and excellent catalytic performance for N2 fixation due to their low work functions and excess electrons. The novel properties of magnetic electrides suggest potential applications in spintronics, topological electronics, electron emission, and as high-performance catalysts. This work marks the beginning of a new era in the identification, investigation, and practical applications of magnetic electrides.

## ■ INTRODUCTION

Electride is a special category of materials, with the features of excess electrons serving as anions which are trapped in interstitial position of crystal structure and not participate in bonding. [1-4] Differ from metals and intermetallic, the excess electrons in electrides are almost localized relating to salts with *F* centers. [5, 6] The reported electrides in early time are organic materials formed by the complexation among cationic alkali metals ions and crown ethers. [7-11] However, most of them are suffered from intricate stability issues, their practical applications were greatly restricted under ambient conditions. [2, 3, 9-11] The first inorganic electride is $12CaO{\cdot}7Al_2O_3$ ($C12A7{:}4e^-$) that was synthesized in 2003, the excess electrons are confined in zero-dimensional (0D) lattice cavities. [12] The high stability at room temperature and rich intriguing characteristics paved the solid foundation for the flourishing research on the physics and applications of inorganic electrides. [13-15] In recent years, a variety of inorganic electrides with excess electrons localized in 1D-linked channels, 2D planes, and 3D spaces has also been revealed. [5, 6, 16-19]

The special distributions of interstitial electrons in electrides render them less tightly bound to the lattice than those around atoms. As the result, electrides possess unique properties such as high carrier concentration, low work function, effectively thermionic electron emission at low temperature, surface magnetism, and high density of active centers. [5, 6, 13, 16-21] These remarkable features endow electrides wide application prospects in electronic emission devices, [22, 23] organic light-emitting diodes, [24, 25] high-efficiency catalysis, [14, 26-29] spintronics, [30-32] rechargeable ion batteries, [33-36] and superconductivity. [15, 37-39] More interestingly, topological electride has been recently developed based on intermetallic $Ca_3Pb$. [40] Topological electrides can realize the enhanced quantum phenomena and ultrahigh electron mobility due to the combined features of electrides and topological electronic states, leading to the weak spin-orbit coupling (SOC) gap and floating surface states as well as various topological states. [31, 39-46, 68, 69, 82]

Owing to the outstanding properties and broad applications, great effort has been paid to explore new electrides. Computational exploration as an economic approach has been widely used to identify electride candidates including material screening from existing crystal database, electride prediction from artificial structure prototypes, and inverse electride design based on functionality-tailored or oxidized counterparts.[19, 20, 42, 47-54] Lots of electride candidates which were identified from theoretical predictions have then been confirmed by the followed experiments.[12, 17, 18, 32, 55-59] The pioneer works demonstrated the power and effectiveness of computational approaches for electride explorations, but there are still several open questions to be further addressed. First, most of the previous studies excluded the materials which contain the transition metals (TMs) due to intricate multiple valences of TMs, leading to the omission of the screening; as a matter of facts, a few materials containing TMs have been experimentally identified as electrides. Second, previous electride screenings were mostly performed based on special crystal structure or material family, which are far from traversing the whole material database. Third, the electride screenings mostly focused on nonmagnetic (NM) materials, the magnetic electrides especially antiferromagnetic (AFM) electrides are extremely scarce. Recently, Zhu *et al.*[51] performed a systematic electride screening based on the Inorganic Crystal Structure Database (ICSD; containing ~30,000 pristine materials),[60] 167 potential electrides are identified. Even though, this electride screening is incomplete since the latest Materials Project database has more than 144,595 materials now.[61] Therefore, a thorough electride screening based on the latest Materials Project database and with detailed magnetic studies is urgently required.

Here, we screened the magnetic electrides from a high-throughput screening based on first principle simulations and the latest materials project database; To avoid possible omissions, the multiple valence states of transition-metals and magnetism as a crucial description are considered for the first time during the screening. 276 electrides (90 newly identified) have been found, 51 of them are magnetic and can be classified as AFM electrides, ferromagnetic (FM) electrides, and interstitial-magnetic (IM) electrides depending on magnetic origins and orderings. From the systematical investigations on the properties, most of them show high spin-polarization degree, different topological states, and relatively low work functions, making them highly promising for applications in spintronics, topological electronics, electron emission devices, and high-performance catalyst promoters.

## ■ RESULTS AND DISCUSSION

### ■ Database Screening Scheme

**Screening Criteria**. The most typical feature of electrides is the presence of excess electrons, namely the total valence of materials is larger than zero. Therefore, we use the total valence of candidate materials as the first criteria during the screening procedure. This is a simple task for materials without transition-metal elements because of their definite valence states, but much more complicated for the compounds containing TM elements. For the later cases, all potential valences will be considered for transition-metal elements in case of screening omission. Taking Mn as an example, it has been reported to carry +2, +3, +4, +5, +6, +7 valence states in compounds, [62-64] we evaluate the total valence of Mn-based materials by taking Mn at highest valence state (+7) to judge the presence of potential excess electrons. Materials with total valence larger than zero are retained for further assessments. Except excess electrons, additional criteria including the localization degree and the position of excess electrons are introduced to confirm the electride phases, like the electron localization function (ELF) and the component of density of states (DOSs). [40, 47, 48, 51] Besides, topological quantum chemistry (TQC) theory will be also used as the indicator to assess electride phases. [65-68] The material will be regarded as an electride only if all the criteria are satisfied. The definition of these indicators will be detailly provided in the section of "examples for magnetic electrides".

Here, with $Ba_3MnN_3$ as a representative example, we show the identification process and magnetic states of the electrides. $Ba_3MnN_3$ has a hexagonal crystal structure within the space group $P6_3/m$ (No. 176), where the *2b* site is vacant without atom occupation, naturally leaving 1D channels for capturing excess electrons in the lattice (see Figure 1A). $Ba_3MnN_3$ potentially contains excess electrons because the Mn element can carry valence states larger than +3, which ensure $Ba_3MnN_3$ passing the first criteria of the electride screening. We then confirm the electride nature of $Ba_3MnN_3$ by considering ELF, DOS and TQC. From the distribution analysis of potential excess electrons (Figure 1B), sizable electronic states at the *2b* Wyckoff site are observed. The interstitial states link together along the 1D lattice channel, with the localized excess electrons as shown by the ELF map (Figure 1C), implying the existence of an electride phase. From the DOS analysis in Figure 1D, it is found that large proportion of DOS is contributed from interstitial electrons. Especially, in the energy range of -0.3 eV to 0.2 eV, the DOS of interstitial electrons is even larger than the sum from atoms in $Ba_3MnN_3$ (see the bottom panel of Figure 1D). This is a typical feature for electrides where the excess electrons in the interstitial sites make the most contributions to the conducting electronic states. The elementary band representations (BRs) from the TQC theory will be studied as the last step, to strictly demonstrate electride phases from the band symmetry view. [66-69, 82] The BRs express the irreducible representations of bands and their attributions of the Wyckoff sites for atoms or located electrons, and the TQC theory builds a direct connection between the band representation in momentum space and the atomic orbital characteristics in real space. For electrides, the BRs of bands around the Fermi level will belong to interstitial Wyckoff sites. Based on the *Irvsp* and *pos2aBR* programs, [70, 71] we obtain that the BRs of the Ba, Mn, and N atoms in $Ba_3MnN_3$ are *A@6h*, *A/E@2c*, and *A@6h*, respectively, see Figure 1E. The bands far away from the Fermi level follow the BRs of atoms. However, the BRs for the three bands near the Fermi level belong to a new BR of $A_{1g}$@*2b* with *2b* the interstitial Wyckoff sites, which unambiguously verify the electride feature of these bands from the symmetry point of view. Following the strict screen process, $Ba_3MnN_3$ is finally confirmed as a 1D electride phase.

We then determine its ground magnetic state of $Ba_3MnN_3$ crystal by performing the detailed DFT calculations. [72] Mn atoms in a honeycomb lattice of $Ba_3MnN_3$ (Figure 2A) contain unfilled *3d*-shells and usually carry magnetic moment, we fully consider five magnetic states including NM, FM, Néel AFM (NAFM), stripe AFM (SAFM), and zigzag AFM (ZAFM), as shown in Figure 2B. NAFM is always the ground state with the lowest energy no matter the chosen of Hubbard U values (0-4 eV) and exchange correlation functionals (including PBE, LDA and PW91) [See Figure S1 in Supporting Information (SI)]. The calculations confirm that $Ba_3MnN_3$ is a typical AFM electride, as shown in Figure 2C.  Similar screening strategies are also applied to identify NdScGe as a FM electride [73] and $Ca_5Bi_3$ as an interstitial magnetism (IM) electride where the magnetism is mostly from interstitial excess electrons (See Figures 2C and D). Besides, the ELFs of the above three electrides under the magnetic and nonmagnetic states are shown Figure S2-I and II in SI. We find the formation of the electride phase in these materials is independent with their magnetic configuration.

**Screening Process and Materials Analysis**. The flowchart of the database screening is shown in Figure 3A. The source materials are from the Materials Project database. [61] It covers the entire Inorganic Crystal Structure Database (ICSD) [60] and includes abundant metastable structures of materials that are promising to be synthesized in experiments. We divide the materials in the database into two categories depending on whether they contain transition-metal elements. Previous screening usually ignores the compounds containing TM elements because of the presences of partially occupied *d* shells and associated multiple valences which were believed to be unfavorable for the formation of electrides. [42, 47, 48] Considering the calculation cost, we here rule out materials with more than four specifies of transition-metal elements and those with over 100 atoms in the primitive unit cell. From the initial screen, the candidate number in the database is reduced into 27130, where 13200 of them contain transition-metal elements.

Totally 276 electrides are eventually screened from the high throughput calculations, where 90 of them are newly revealed. The full list of electrides is provided in Table SI-III in SI. Besides, the corresponding ELF maps of electrides are also provided (see Figures S3-S4 in SI). As classified by the dimensionality of the cavities confining excess electrons (Figure 3B), 72.10% (199 specifies) are 0D electrides, 20.67%, 4.43% and 2.89% are 1D, 2D and 3D electrides, respectively. The number of high dimensional electrides sharply decreases due to the more stringent formation conditions (specific analysis can be found in the sections 7 and 8 (Figures S5 and S6 in SI]. From the compositions of electrides, more than 60% (175 species) are binary electrides (e.g., $Ti_3Br$, $Sc_5Sn_3$, $Lu_5Rh_3$), which are composed by one main-group element and one transition-metal element (See Table SII in SI). 99 ternary electrides (e.g., CeScSb famlily, and $Gd_2MgNi_2$ family) and 2 quaternary electrides [$K_4Al_3(SiO_4)_3$ and $La_8Sr_2(SiO_4)_6$] are identified (see Figure 3C). The number of electrides is significantly reduced with an increasing variety of the elements, indicating the reasonability to ignore the compounds with more over 4 elements during the screening process as above. We also notice that over 74.7% (206 species) of the identified electrides have been synthesized based on available experiments (See Table SI in SI), 70 species have similar structures with existing materials, to exhibit excellent dynamical stability. By examining the magnetic properties, we found that 225 species are NM, 51 species are magnetic, including 12 AFM, 13 FM and 26 IM electrides (see Figure 3D). The number of IM electrides plays the dominant role, indicating that excess electrons in the interstitial sites are the important origins of the magnetism in electrides especially those without rare-earth/transition-metal elements. The full list of the identified magnetic electrides can be found in Table I, where 42 species are reported for the first time.

In the following, we will explore the unique properties and potential applications of magnetic electrides, including the magnetic states, the topological states, work functions, and catalytic performances.

## ■ Physical and Chemistry Properties

**Magnetic Properties.** In contrast to the conventional electrides, the magnetic electrides in the work (39 FM and IM electrides, see Table I) are featured with high spin-polarization (*P*) and various magnetic ground states. Three typical magnetic electronic states including traditional magnetic metal, magnetic semiconductor, and half metal have been found (see Figures 4A and B). 84.3% (43 species) are magnetic metals, 7.84% [4 species, namely $Ca_5Sb_3$ (*Pnma*), $Sr_5Sb_3$ (*Pnma*), $LaBr_2$ ($P6_3/mmc$), $K_2RbPt$ (*P4/nmm*)] are magnetic semiconductors, and 7.84 % [4 species, namely $Ca_5Bi_3$ (*pnma*), $Sr_5Bi_3$ (*pnma*), $Ba_5As_3$ ($P6_3/mcm$), $Ba_5Sb_3$ ($P6_3/mcm$)] are half-metals. 20 electrides host high spin-polarization degree larger than 50%, where 4 IM electrides have the 100% spin-polarization (see Figure S7 of SI).

Different from the traditional magnetic materials, two special characteristics have been identified for magnetic electrides. First, the conducting electrons in magnetic electrides are raised from loosely-bounded excess electrons rather than the atomic orbital electrons. Such excess electrons have been well evidenced to show high carrier mobility and low work function. [5, 17] Considering the fact that high carrier mobility and low work function are two crucial elements for the applications of spintronics including spin injection and various spintronics technologies such as spin wave, giant magneto resistance, magnetic tunnel junction, and ferroelectric tunnel junction, [74-77] magnetic electrides which combine high spin-polarization, high mobility and low work function of conducting electrons will undoubtedly show special advantages in these applications comparing to conventional magnets. Second, 25 specifies of identified electrides are IM electrides, namely the magnetism in these materials is merely contributed by the excess electrons in the interstitial lattice sites rather than by the magnetic atoms in conventional magnets (see Figure 2E). Such novel characteristic in IM electrides is expected to serve as an ideal platform to investigate the inherent magnetism of pure electron systems experimentally, [78] which has been a long-standing problem for understanding the fundamental mechanism of magnetism in condensed matters. [79-81]

In Figures 4C-E, we respectively discuss the electronic/magnetic properties of the selected typical examples of magnetic metals, magnetic semiconductors, and half-metals in electrides. CaF is a magnetic metal as shown in Figure 4C, the ELF & SDD maps suggest it is a 0D IM electride with excess electrons locating at the interstitial space of lattice. The spin-resolved band structure and DOSs for CaF indicate that both spin channels have metallic bands crossing the Fermi level but with different densities, leading to a sizable (79.3%) spin-polarization degree. $LaBr_2$ is an example of magnetic semiconductors (Figure 4D), the ELF and SDD maps suggest $LaBr_2$ is a 0D IM electride. Further electronic analysis indicates that it is a magnetic semiconductor with different band gaps for two spin channels around the Fermi level. $Ba_5As_3$ is a half-metal (Figure 4E) and 0D electride with IM magnetism, which is metallic in the spin-up channel but exhibits an insulating gap of 0.298 eV in the spin-down channel.

It is interesting to notice that the magnetic states can be transferred to each other in electrides under different conditions. For example, a 10% lattice compression in $LaBr_2$ will shift the valence band in the spin-down channel above the Fermi level while the band gap in the spin-up channel is retained (see Figure S8 in SI), the semiconducting $LaBr_2$ has transformed into a half metal phase, while the electride character in $LaBr_2$ can be retained. Since most electrides are susceptible to oxidation and hydrogenation because of the presence of interstitial electrons, we here also check the effects of oxidation/hydrogenation on the magnetic states of electrides. Taking $Gd_2C$ (FM metal with an 87% spin-polarization) as an example, it can be transformed into a half metal if one side of $Gd_2C$ is hydrogenated, the spin up bands are metallic but spin-down channels are insulating (see Figure S8 of SI). The half hydrogenated $Gd_2C$ is still in electride phase because the excess electrons are reserved in the interstitial sites. Our works have provided feasible approaches to realize half-metallic electrides by electronic engineering from lattice strain, and hydrogenation/oxidation.

**Topological Electrides**. Topological electrides refer to the electrides with nontrivial band topology, and the outstanding properties of both topological and electride materials. They could show exotic properties such as ultrahigh electron mobility, weak SOC phenomenon, and floating surface states. [40, 42, 43, 82] In topological magnetic electrides, additional novel properties can be expected as a result of the time-reversal symmetry breaking. In this work, all the magnetic electrides identified here are found to show nontrivial band topology with a variety of topological states, as shown in Table I.

Figure 5A provides a brief illustration of different topological states and their nontrivial surface signatures, including abnormal quantum Hall effects (AQHEs), Weyl point (WP), triply-degenerated nodal point (TDNP), Dirac point (DP), nodal line, and nodal surface. 72.5% electrides (37 species) show WPs, 52.9 % (27 species) for NLs, 35.3% (18 species) for NSs, 19.6% (10 species) for AQHEs, 29.4% (15 species) for TDNPs, and 13.7% (7 species) for DPs (See Figure S9 of SI). Among them, WP has the highest ratio since the time-reversal symmetry in magnetic electrides is naturally broken, it well satisfies the formation of WPs from the symmetry point of view. It is worthy to point out some electrides host multiple types of topological states (see Table I). Thus, magnetic topological electrides can provide excellent platforms to investigate potential relevance among nontrivial topological states, excess electrons, and magnetism.

The topological band structures for all the identified magnetic electrides have been investigated and summarized in Figures S10 and S11. Here we take the AFM electride $Ba_3MnN_3$ as an example for specific discussion. As shown in Figure 5B, $Ba_3MnN_3$ features a band crossing along the Γ-A path near the Fermi level. A careful examination on the band degeneracy finds the crossing point is a TDNP (see the 3D band plotting of the inset), formed by a doubly-degenerated band and a singlet. The electronic contribution of the TDNP fermion is almost from the excess electrons, as shown by the PED map in Figure 5C. In addition to the TDNP, the doubly-degenerated band forms a quadratic NL along the whole Γ-A path. Therefore, electride $Ba_3MnN_3$ hosts both TDNP and NL fermions (see Figure 5C). These topological fermions feature nontrivial surface states. As shown in Figure 5D, we can observe clear surface bands originating from both TDNPs and QNLs. The inclusion of spin-orbit coupling (SOC) will gap the QNL and the TDNP pair transforms into DPs (see Figure 5E). Figure 5F shows the surface states for the DPs, where two pieces of Fermi arcs can be observed. These results demonstrate $Ba_3MnN_3$ is an excellent topological electride material to host multiple topological fermions under the AFM state.

Comparing with NM counterparts, the magnetism in magnetic topological electrides can act as a magnetic field (an additional degree of freedom) to regulate the topological states. Taking the IM electride CaF as an example, it has two band crossings near the Fermi level with different band slopes ($P_1$ and $P_2$, see Figure 6A) when SOC is not considered, where $P_1$ is a critical-type formed by a dispersive band and a nearly flat band while $P_2$ is a type-II crossing with a completely titled Weyl cone (the comparison among different types of Weyl fermions is shown in Figure S12 in SI). Our further calculations show these band crossings locate on a NL, centering the Γ point in the $k_z$ =0 plane. Considering different magnetization direction, the NL can be transformed into different topological states under SOC. We have evaluated the magnetocrystalline anisotropy of electride CaF, and find that the in-plane magnetizations have lower energies than that of out-of-plane. Remarkably, the energy difference for in-plane directions is lower than 0.1 meV, indicating the magnetic configuration in can be readily regulated by external magnetic field. With this in mind, we investigate the topological band structure of CaF under different magnetizations. Typically, under the in-plane [010] magnetization direction, $P_2$ is gapped while $P_1$ is not affected, leaving a pair of critical WPs (see Figure 6B and 6C). This transition can be understood from the symmetry point of view. Under the [100] magnetization, the mirror symmetry $M_z$ which protects the nodal line is broken. However, the $M_{(010)}$ symmetry in the *M*-*Γ*-*M'* path still retains, which will allow band crossing along this path. If the magnetization direction is shifted to the [110] direction, the critical WPs $P_1$ will be gapped, but the type-II Weyl points $P_2$ in the *K*-*Γ* path are preserved under the protection of the $M_{(-100)}$ symmetry. Under the out-of-plane [001] magnetization, the whole NL can be protected because the $M_{(001)}$ symmetry is not broken. The magnetic electride CaF can switch among rich topological states as regulated by external magnetic fields, including the critical Weyl state under [010] direction, type-II Weyl state under [110] direction, and NL state under [001] direction, as illustrated in Figure 6C. Thus, magnetic topological electrides are good platforms to investigate variable topological states and corresponding topological phase transition in a single material.

## ■ Work Function and Catalytic Applications for Magnetic Electrides

**Work Function**. Electrides are attractive as electron donors for various applications like electron emitters due to low work functions. [5, 6, 13, 17, 20] Figure 7A schematically illustrates the mechanism of an electron emission electrode based on an electride, where the excess electrons in electride can be stimulated out readily by the electrical field because of their low work function. Since most of the magnetic electrides are identified for the first time in this work, we here systematically evaluate the work functions of the magnetic electrides. The results are summarized in Table I and Figures S13 -S15 of SI. We find that more than 60.7% (31 specifies) of the magnetic electrides possess work function lower than 4 eV, which are lower than the values of typical electron emission materials including transition metals Ni (4.6 eV), Cu (4.65) Cd (4.07) Pd (5.12 eV), and graphene (4.6 eV). [83, 84] As previously reported, among known electrides, nonmagnetic $Ca_2N$ shows extremely low work function (~3.4 eV). [17] In the work, we find 27 magnetic electrides even show lower work function than that of $Ca_2N$ (see Figures S13 and S14 of SI). Especially, the work function of AFM electride $Ba_4Al_5$ is only 2.02 eV, which is the lowest among materials reported so far. These results suggest magnetic electrides are highly promising for electron-emitting applications with quite low work functions.

We further verify that the low work functions in these electrides arise from the excess electrons at the interstitial space of lattice. We here take the magnetic electride CaF as an example. The crystal structure of CaF is shown in Figure 7B, where the Ca and F atoms are layered in the honeycomb lattice. A primitive cell contains two formulas of CaF, thus hosts two excess electrons (represented as CaF:2$e^-$ in the following). The ELF map in the inset of Figure 7B clearly shows the presence of excess electrons in CaF:2$e^-$ locating at the interstitial sites. Then we gradually annihilate these excess electrons by doping different degrees of holes in the system and investigate the impacts on the work function. The ELF maps for doping 0.5 holes (CaF:1.5$e^-$), 1 hole (CaF:1$e^-$), 1.5 holes (CaF:0.5$e^-$), and 2 holes (CaF:0$e^-$) are shown in the insets of Figure 7B. We can find that, with the increasing of degree of hole-doping, the excess electrons are indeed gradually annihilated, where the case of CaF:0$e^-$ would have none excess electrons. The work functions for different cases are shown in Figure 7B. One can observe that the primitive CaF:2$e^-$ which possesses two excess electrons shows the lowest work function. However, when the excess electrons are gradually annihilated, the work function increases correspondingly, indicating the low work function in electride is highly related to the excess electrons.

In addition, we find the work functions in electrides show obvious anisotropy on different surfaces, which shows high correspondence with the distribution of excess electrons. We take the 1D electride $Ba_3MnN_3$ as an example. The insets of Figure 7C show the ELF maps of $Ba_3MnN_3$ from different sides of views. We evaluate the emission condition of excess electrons from different surfaces. As shown Figure 7C, (001) surface can provide a large 1D channel for the emission of excess electrons (see the circled region), along which the electron emission is almost immune from the interaction with the bonds or atoms. As the result, low work function can be expected on this surface. For the (110) surface, it also hosts sizable channels for the electron emission (see the framed regions in Figure 7C). However, the presence of atomic layers between channels would partly confine the emission of excess electrons. Thus, the work function on the (110) surface should be higher than that on the (001) surface. For other surfaces such as (111), there unfortunately exist no large area of emission channels for excess electron (see the bottom panel of Figure 7C), which will lead to high work function on this surface. With this in mind, we compare the work function for these surfaces and the results are shown in Figure 7C. We find the (001) surface (2.73 eV) indeed shows the lowest work function among these surfaces. (110) surface (2.86 eV) possesses a slightly higher work function than the (001) surface, while that on the (111) surface (4.68 eV) is greatly higher than the (110) and (001) surfaces. This trend of work function exhibits a good accordance with the diverse conditions of emission channels for excess electrons shown in the insets of Figure 7C. Similar anisotropy of work functions depending on the surfaces and distributions of excess electrons can also be found in 2D electrides (see Figure S16 in SI). These results fully support that the presence and distribution of excess electrons are responsible for the low work function in electide. Besides, the clarification of such mechanism can also provide guidance for designing materials with low work functions based on electrides.

**Catalytic Applications.** Besides the applications as electron emitters, electrides have also been proven as high-performance catalyst promoters for activate various molecules including $H_2$, $N_2$, $NH_3$, CO, $CO_2$, etc. [5] Among them, electride-based catalysts have been the most widely investigated under reductive conditions such as ammonia ($NH_3$) synthesis and selective hydrogenation reactions [such as hydrogen evolution reaction (HER)]. We have evaluated the Gibbs free energy for HER ($\Delta G_{H*}$) in the magnetic electrides. We find most of them show high $|\Delta G_{H*}|$ values (> 0.6 eV). This arises from that excess electrons (especially those locating near the surface) tend to show a strong combination with H atom, leading to a highly-negative value of adsorption energy. Even though, we still identify two magnetic electrides showing low $|\Delta G_{H*}|$ (0.22 eV for $Ba_4Al_5$, and 0.28 eV for CaF), indicating them highly potential as HER catalysis. For the electrocatalytic ammonia synthesis, several electrides have been evidenced as high-performance catalyst promoters when they are loaded with the transition-metals (transition-metals/electirdes such as Ru/C12A7:4e$^-$). [14, 85, 86-88] Two factors were believed to promote the catalytic performance in electrides as shown in Figure 8A: first, the low work function from excess electrons endows high electron donation power of electrides, which can provide a large amount of electrons to occupy the antibonding π-orbital of $N_2$, resulting in low apparent activation energy during the ammonia synthesis; [28] second, the large interstitial space in the electride lattice allows reversible hydrogen storage and release in the reaction, which can effectively prevent hydrogen poisoning on catalyst surface and promote the $NH_3$ synthesis rate. [5, 28, 29, 88] Among known electrides, C12A7:4e$^-$ is the most used catalyst promoters, due to its low work function (2.4 eV) and good water durability. Although TM/C12A7:4e$^-$ show excellent catalysis performance, the aggregation of the transition-metal nanoparticles remains as a challenge during the TM/C12A7:4e$^-$ synthesis because C12A7:4e$^-$ itself contains no transition-metal or rare-earth elements. [5] Therefore, much effort has been paid to explore desirable transition-metal/rare-earth-containing electrides. In our electride screening, we have totally identified 83 species of such electrides. Among the magnetic electrides shown in Table I, 7 electrides contain transition-metal/rare-earth elements. Especially, several can show comparable work functions with C12A7 like $Ba_3MnN_3$ (2.45 eV), $Sr_3MnN_3$ (2.59 eV), $Gd_2MgNi_2$ (2.66 eV), $Dy_3Co$ (2.73 eV). We also evaluate the work functions of the NM electrides. 17 specifies of NM electrides show the work functions lower than 3 eV. These electrides are promising candidates to replace C12A7:4e$^-$ as high-performance catalyst promoters (see Figures S15 and S17 of SI).

Taking $Gd_2MgNi_2$ as an example, we here prove the feasibility to use the electrides as catalyst promoters. Following the well-known Ru/C12A7:4e$^-$, Ru/$Gd_2MgNi_2$ is selected for the ammonia synthesis. We first evaluate the promotion of $N_2$ cleavage process of electride $Gd_2MgNi_2$. When the bare Ru is used for $N_2$ fixation, the apparent activation energy barriers for the initial and final state with respect to the intermediate state are up to 4.31 eV and 5.29 eV, respectively [see the (i) of Figure 8B]. However, once Ru is loaded on $Gd_2MgNi_2$, the two barriers are sharply reduced to 1.17 eV and 1.65 eV [see the (ii) of Figure 8B], which are close to that of Ru/C12A7 (*83*). These results suggest the presence of electride $Gd_2MgNi_2$ can strongly promote the $N_2$ fixation during the ammonia synthesis. The detailed processes of $N_2$ fixation in Ru and $Gd_2MgNi_2$/Ru are shown in Figures S18-21 of SI. In addition, the presence of large interstitial lattice space in electrides is also believed to play an important role in the storage and release of hydrogen which can effectively prevent hydrogen poisoning on catalyst surface and promote the $NH_3$ synthesis rate. [5, 6, 28, 29, 88] For this process, the diffusion barrier of hydrogen is an important indicator. The adsorption energy of H at the interstitial site *1b* (0.5, 0.5, 0.5) in $Gd_2MgNi_2$ is calculated to be -0.75 eV, indicating the feasibility of hydrogen storage. We also evaluate the barrier of hydrogen diffusion from the interstitial site to the surface of $Gd_2MgNi_2$ as 1.98 eV [see Figure 8C], which is significantly lower than that of C12A7 (3.1 eV). [86] More inspiringly, the diffusion barrier of hydrogen can be even lower in several 1D magnetic electrides identified in this work, such as $Ba_3MnN_3$ [0.9 eV, see Figure S22 of SI]. Therefore, by simultaneously promoting the $N_2$ fixation and $H_2$ storage & release processes, the electrides identified in this work will be promising to act as supporting materials for catalysis.

## ■ CONCLUSIONS

In conclusion, from high-throughput calculations, we screened electrides based on the latest Materials Project database. Totally 276 electrides are found, and 90 of them are reported for the first time. Remarkably, 51 electrides are identified to show magnetic ordering as the ground state, including 12 AFM electrides, 13 FM electrides, and 26 IM electrides. The electrides can exhibit as magnetic metals, magnetic semiconductors, and also half metals with the 100% spin-polarization. They also show a variety of topological states. By combining electride feature and the magnetic freedom, they can provide excellent platforms to investigate the excess-electrons-induced nontrivial fermions and tunable topological phase transitions in electrides. In addition, due to the low work functions arising from the presences of excess electrons, they are highly promising for applications as electron-emitting and high-performance catalyst promoters. Our work not only significantly enriches electride candidates especially the magnetic ones, but also provides guidance for their applications in various fields.

## ■ ASSOCIATED CONTENT

**Supporting Information**

Supporting Information contains "Calculation methods", "Summary of 276 electrides", "ELF maps of electrides", "Work function", "Electronic band structures", "Cleavage process of $N_2$", "Hydrogen transport" etc.

## ■ ACKNOWLEDGEMENTS

This work is supported by the National Natural Science Foundation of China (No. 12274112). The work is funded by the Science and Technology Project of the Hebei Education Department, the Nature Science Foundation of Hebei Province, the S&T Program of Hebei (No. A2021202012), and the fund from the State Key Laboratory of Reliability and Intelligence of Electrical Equipment of Hebei University of Technology (No. EERI_PI2020009).

## ■ REFERENCES

## Figures and Captions

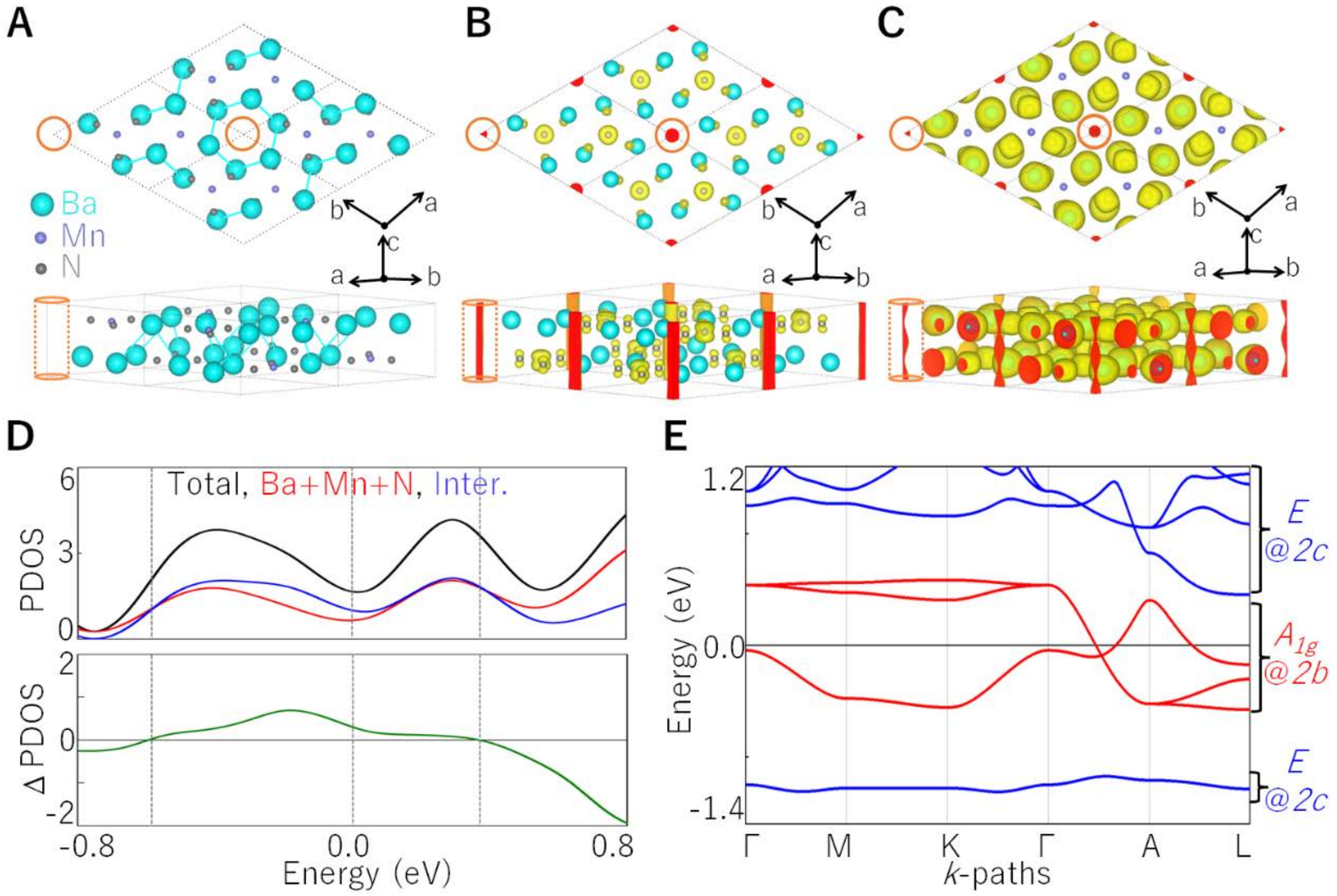

**Figure 1. The judgment indicators for electrides.** (**A**) The crystal structure of 2×2×1 supercell for electride $Ba_3MnN_3$. The circled regions show the interstitial sites of the lattice. (**B**) The partial electronic density (PED) and (**C**) electron localization function (ELF) of electride $Ba_3MnN_3$. The isosurface values for ELF and PED maps are taken as 0.55 and 0.0003 e/Å$^3$, respectively. The energy range of PED is -0.5 eV < $E$-$E_F$ < 0.5 eV. In (**B** and **C**), the circled 1D channels show excess electrons are localized at the *2b* interstitial site of electride $Ba_3MnN_3$. (**D**) The top region shows the partial density of state (PDOS) for electride $Ba_3MnN_3$; the bottom region shows the difference value of DOSs between orbital electrons from atoms and excess electrons. (**E**) The electronic band structure of electride $Ba_3MnN_3$. The bands near the Fermi level (-0.4 eV ~ 0.4 eV) are contributed by the excess electrons of *2b* interstitial site, and the bands away from the Fermi level are contributed by the *d*-orbital of the Mn atom in *2c* site. *2c* and *2b* represent the Wyckoff sites of Mn atoms and interstitial positions. $E$ and $A_{1g}$ represent the irreducible representation (IR) of the bands.

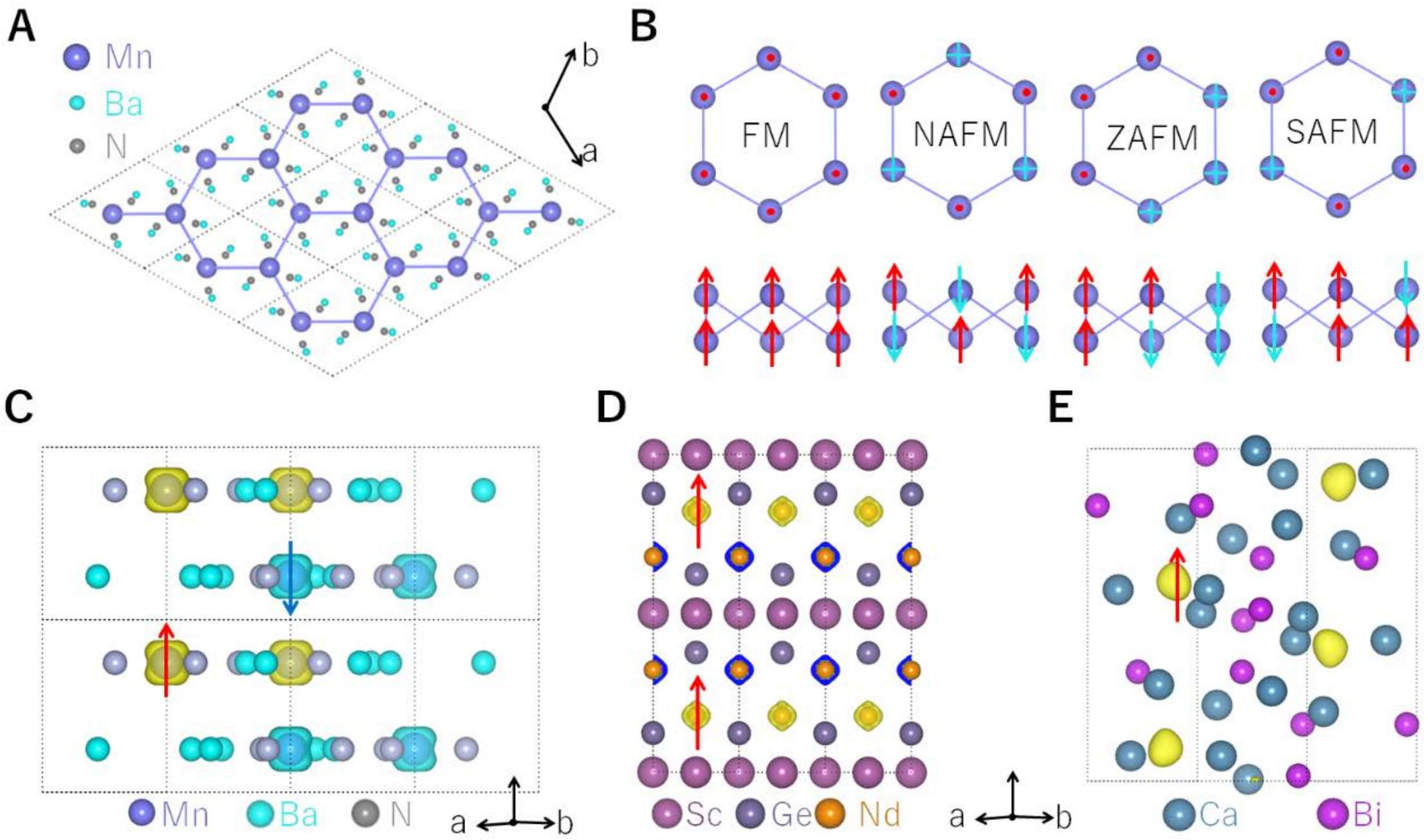

**Figure 2. The determination of magnetic ground states of electrides.** (**A**) The top view of 3×3×1 supercell structure for electride $Ba_3MnN_3$. The magnetic Mn atoms form the typical hexagonal structures. (**B**) Based on the hexagonal structure of Mn atoms in (a), the magnetic ground states of electride $Ba_3MnN_3$ are classfied into four categories: ferromagnetic (FM), Néel antiferromagnetic (NAFM), stripe antiferromagnetic (SAFM), and zigzag antiferromagnetic (ZAFM), respectively. (**C-E**) show the spin differential density (SDD) maps for AFM electride $Ba_3MnN_3$, FM electride NdScGe, and IM electride $Ca_5Bi_3$, where yellow and blue charges represent spin up and spin down charges.

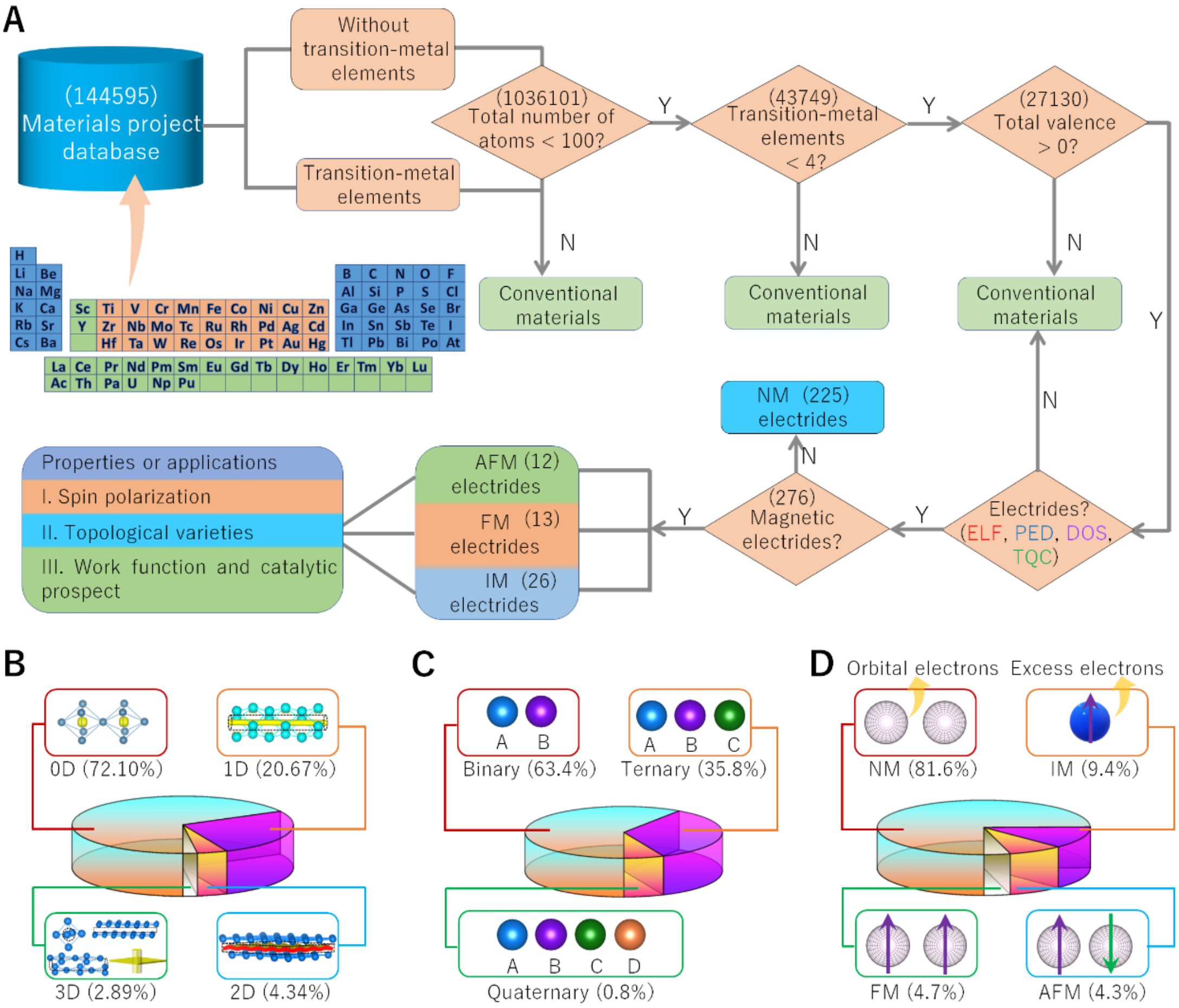

**Figure 3. Electrides screening flow chart and the classification of electrides.** (**A**) Flow chart for searching potential electrides in materials project database (144595 species) by high-throughput screening. Among them, the periodic table of the elements is classified as three groups: the main group elements (blue), the transition group elements (orange), and the lanthanide elements (green). ELF, PED, DOS, and TQC represent electron localization function, partial electronic density, density of states, and topological quantum chemistry theory, respectively. Besides, NM, AFM, FM, IM represent non-magnetic, antiferromagnetic, ferromagnetic, interstitial-magnetic, respectively. The screened electrides are systematically classified by (**B**) the dimensions of excess electrons [namely 0D (cage), 1D (channel), 2D (interlayer), 3D (stereoscopic)], (**C**) stoichiometry (namely binary, ternary, quaternary), and (**D**) magnetism (namely NM, FM, AFM, and IM). In (b), the 0D, 1D, 2D and 3D examples of electrides in the insets are $Ca_3Pb$, $Sr_3CrN_3$, $Sr_2N$, and $K_3O$, respectively (see **Figure S4** of **SI**).

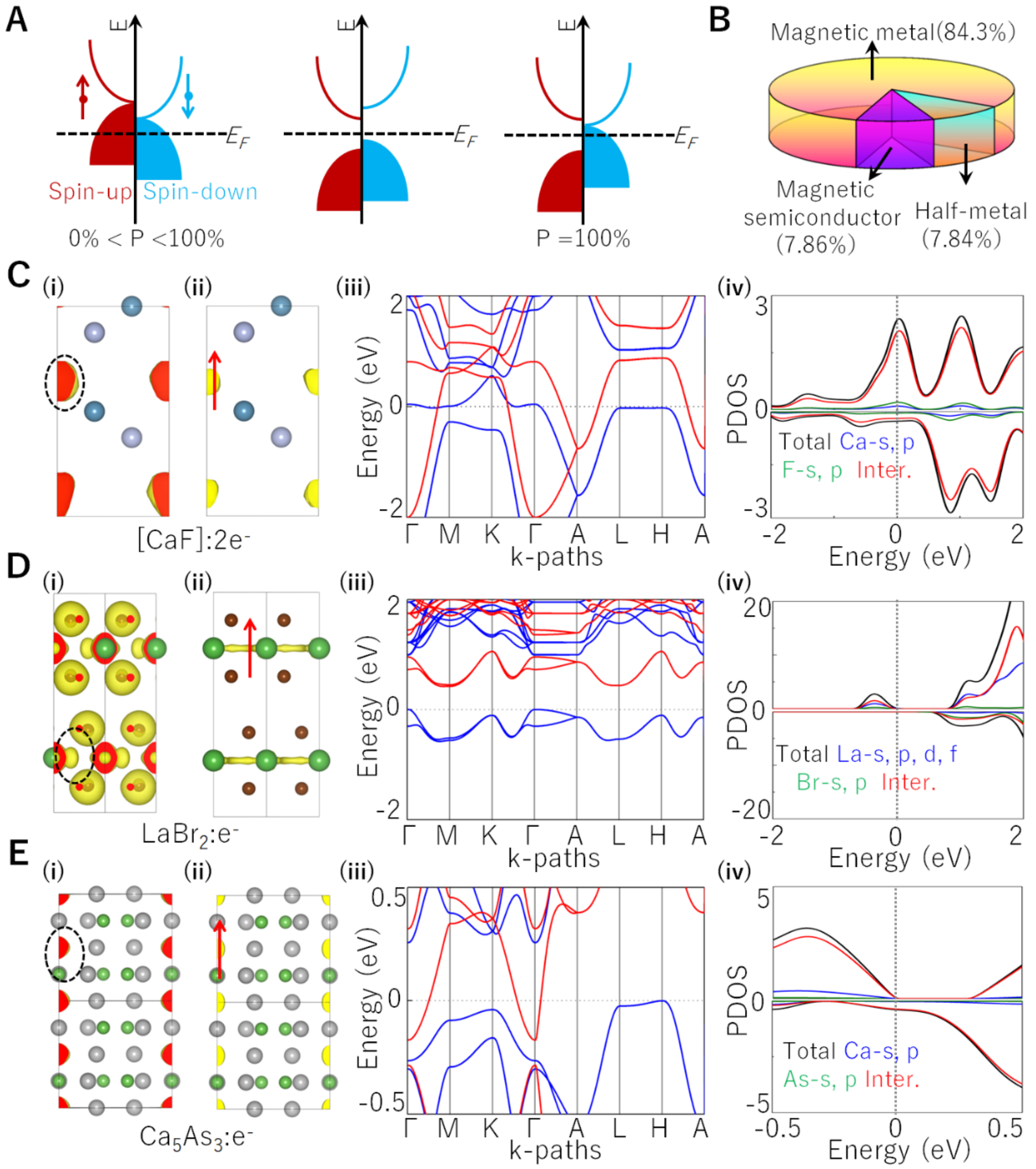

**Figure 4. Magnetic electride examples for metal, semiconductor, and half-metal.** (**A**) The schematic diagram of band structure magnetic metal, semiconductor, and half metal. (**B**) The proportion of magnetic metal, semiconductor, and half metal in identified magnetic electrides. (**C**) The ELF (i) and SDD (ii) maps for electride CaF. (iii) and (iv) are the electronic band structures and PDOSs of electride CaF. In the band structure, the bands for spin up and spin down channels are given as blue and red lines. (**D**) and (**E**) are similar with (**C**) but for electrides $LaBr_2$ and $Ca_5As_3$, respectively. The isosurface value for ELF in (C), (D), (E) is chosen as 0.6. The isosurface value for SDD in (C), (D), (E) is chosen as 0.005, 0.009, 0.008 $Bhor^{-3}$, respectively.

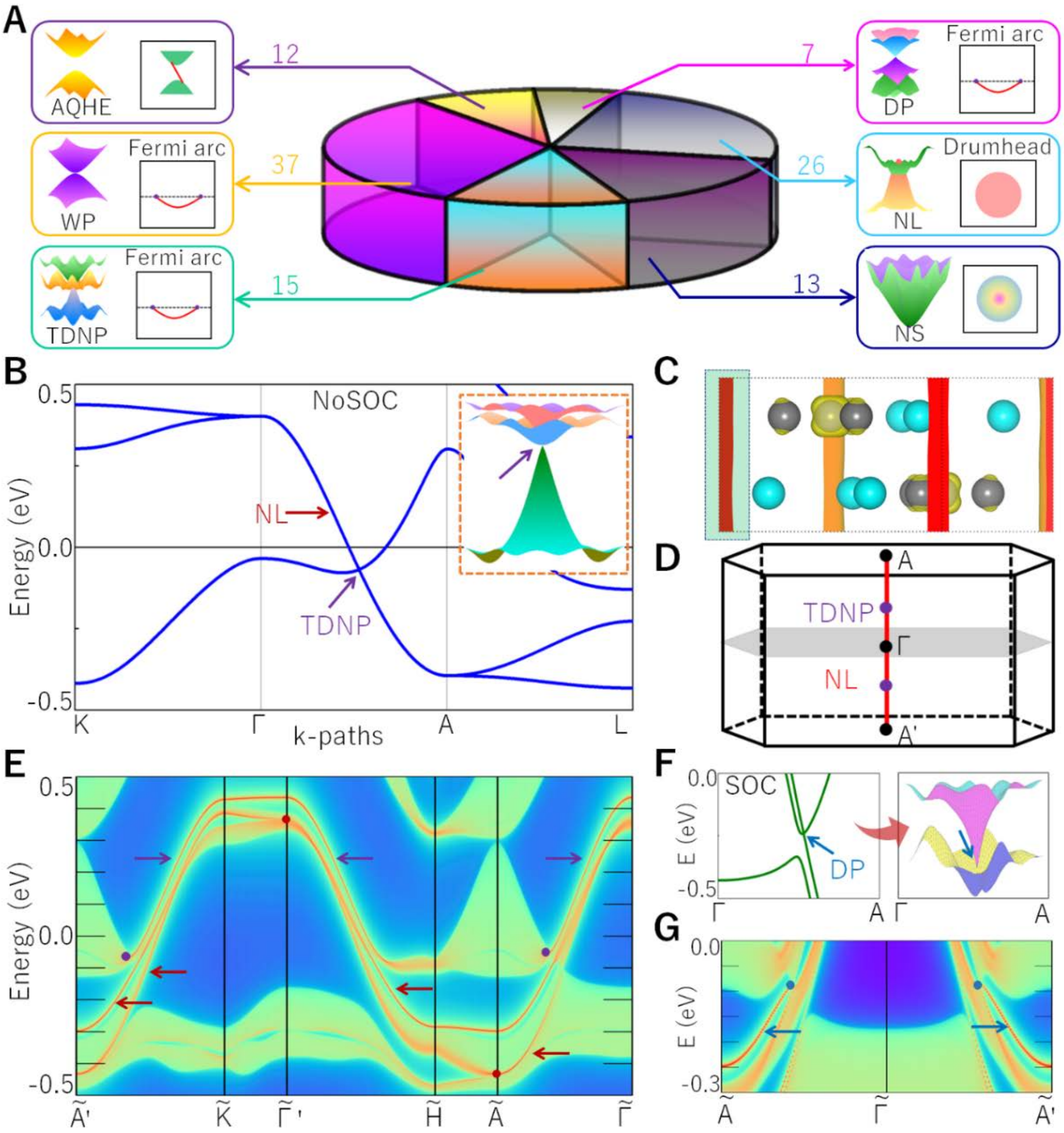

**Figure 5. Classification of magnetic topological electrides based on their topological states.** (**A**) Classification of topological electrides with anomalous quantum Hall effect (AQHE), Weyl point (WP), Dirac point (DP), triple-degenerate nodal point (TDNP), nodal line (NL), and nodal surface (NS). Illustrations for these topological states and their surface states are also provided. (**B**) The electronic band structure without SOC for $Ba_3MnN_3$. The inset shows the 3D dispersion band of TDNP. (**C**) The PED map of $Ba_3MnN_3$, where the energy range of PED is taken within -0.5 eV < E-E$_F$ < 0.5 eV. The isosurface value for PED is chosen as 0.0035 Bohr$^{-3}$. (**D**) The position of the TDNP and NL in the Brillouin zone (BZ) of $Ba_3MnN_3$. (**E**) The surface state on (100) surface without SOC for $Ba_3MnN_3$. The purple and red dots represent the positions of NL and TDNP, respectively. The arrows point to Fermi arcs and drumhead surface states. (**F**) The electronic band structure of $Ba_3MnN_3$ under SOC, and the corresponding 3D band dispersion of Dirac point. (**G**) The surface state on (100) surface with SOC included for $Ba_3MnN_3$. The blue dots represent the positions of DP. The arrows point to Fermi arcs surface states.

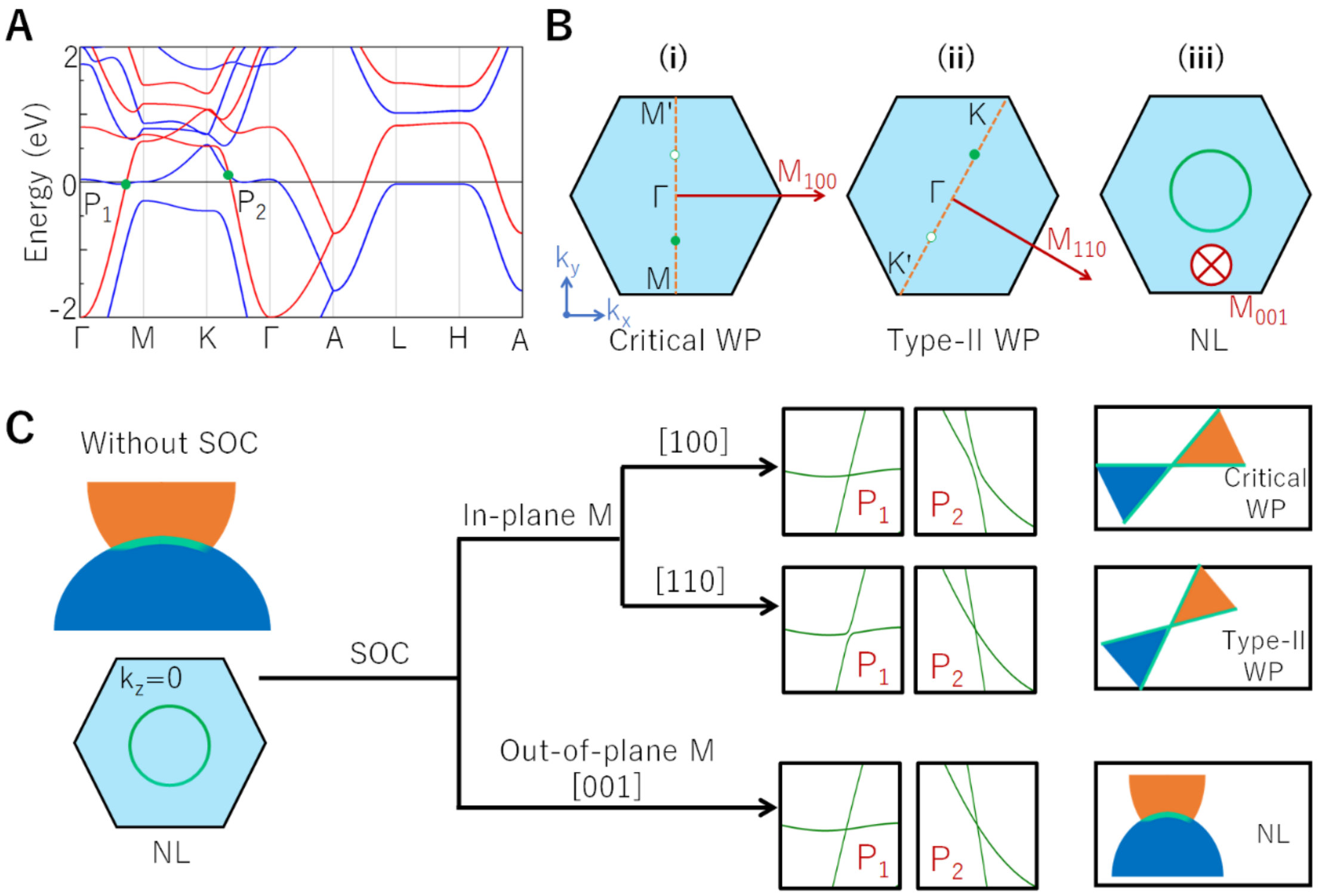

**Figure 6. Topological phase transition of magnetic electrides.** (**A**) The electronic band structure of IM electride CaF. The crossing points near the Fermi level are denoted as $P_1$ and $P_2$, respectively. (**B**) Illustration of topological states under different magnetization directions. (i), (ii) and (iii) for the [100], [110], [001] directions, respectively. (**C**) Schematic diagram of different topological phases realized in the electride CaF, which shows NL without SOC, and critical WP, type-II WP and NL under SOC with different magnetization applied. The band structures near $P_1$ and $P_2$ for these cases are also provided.

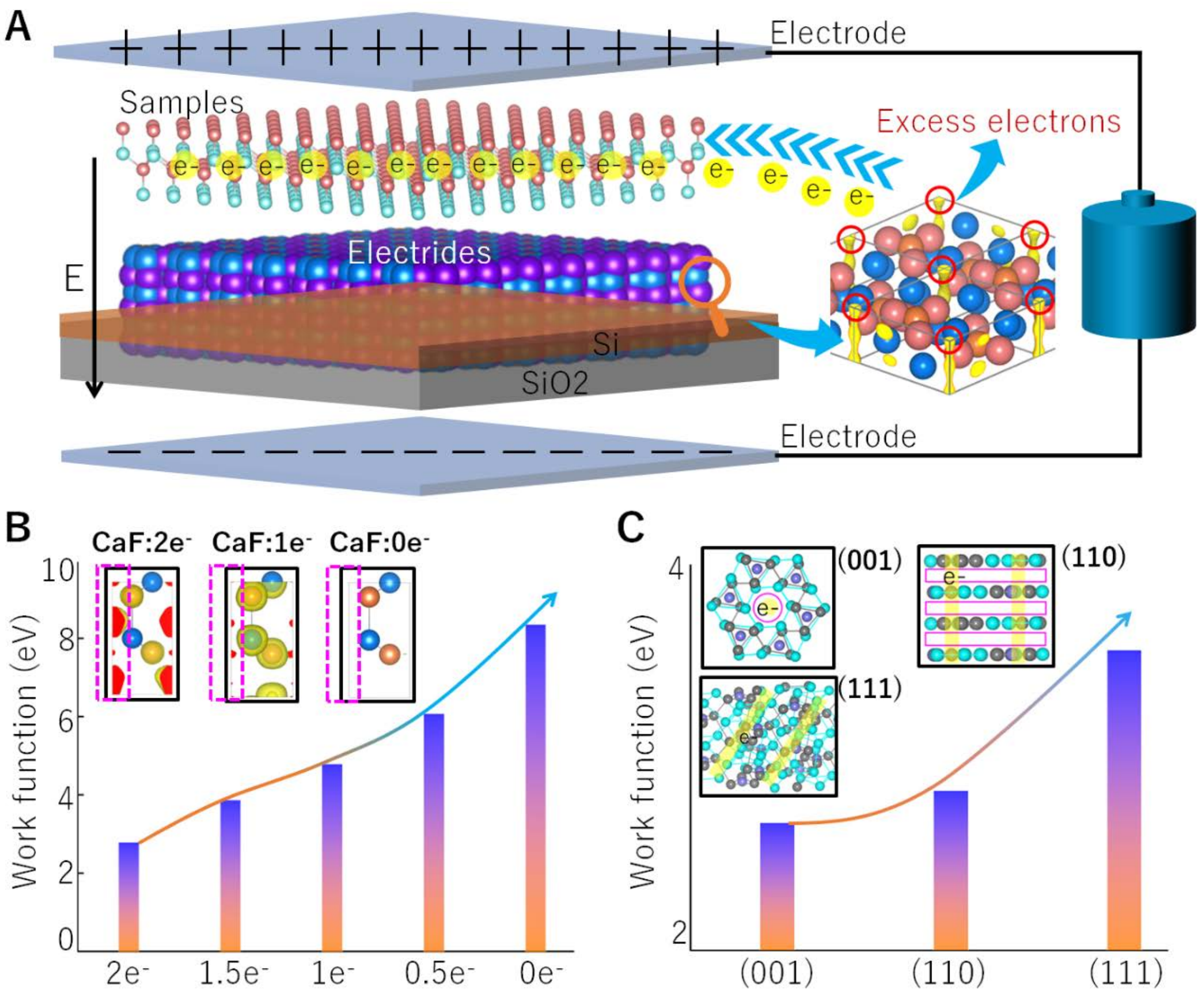

**Figure 7. Work functions of magnetic electrides.** (**A**) Schematic diagram of the mechanism of an electron emitter. Working principle: by epitaxial growth method, the desired materials (samples) are grown on a specific surface of the electrides [e.g. (001) surface]; then an effective electric field is appled to inject excess electrons into the desired samples. (**B**) The work functions for electride CaF with excess electrons gradually annihilated by hole doping. CaF:1.5$e^-$, CaF:1$e^-$, CaF:0.5$e^-$, and CaF:0$e^-$ are for doping 0.5, 1, 1.5, and 2 holes. The insets show the ELF maps for different cases. The isosurfaces for ELF maps are taken as 0.7. (**C**) The work functions of electride $Ba_3MnN_3$ on different surfaces including (001), (110) and (111). The insets are the crystal structure of electride $Ba_3MnN_3$ from [001], [110] and [111] view of sight. The yellow charge shows the location of the excess electrons.

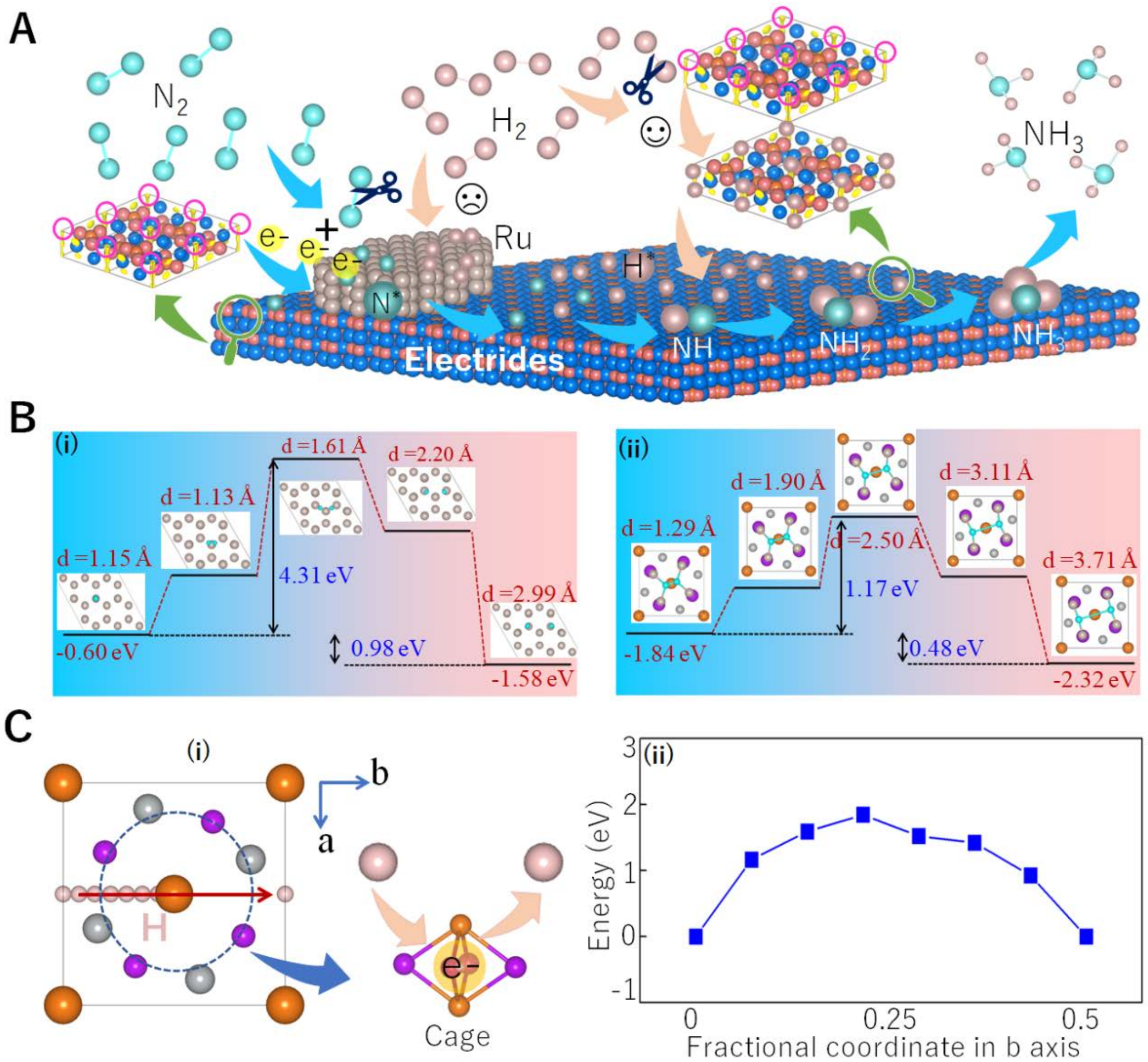

**Figure 8. Catalytic performance of magnetic electrides** (**A**) Schematic diagram of catalytic ammonia synthesis of electrides loaded Ru (electrides/Ru). The catalytic promotion of electrides works in two processes. First, the low work function of excess electrons endows high electron donation power of electrides, which can provide a large amount of electrons to occupy the antibonding π-orbital of $N_2$ and low apparent activation energy during the ammonia synthesis. Second, the large interstitial space in the electride lattice allows reversible hydrogen storage and release in the reaction, which can effectively prevent hydrogen poisoning on catalyst surface and promote the $NH_3$ synthesis rate. (**B**) Energy profiles for $N_2$ activation on (i) Ru and (ii) Ru/$Gd_2MgNi_2$, where "*d*" represents the bond length of N-N. (**C**) The left plane of (i) shows the H diffusion along the *b* axis in lattice of electride $Gd_2MgNi_2$. The right plane of (i) is the schematic diagram of the reaction between H atom and excess electrons within the 0D cage. (ii) The energy barrier for H diffusion between the (0.5, 0, 0.5) and (0.5, 0.5, 0.5) internal coordinates shown in (i).

**Table I** Space groups, dimensions of excess electrons, magnetic ground states, spin polarization degree (*P*), topological states and work functions ($\Phi_{WF}$) of the identified 51 magnetic electrides in this work.

| Serial number | Space group | Electrides | Dimensions | Magnetic group state | *P* (%) | Topological states | WF ($\Phi_W$) |
|---|---|---|---|---|---|---|---|
| 1 | pnma | $Dy_3Co$ | 0-D | AFM | 0 | WP, NL, DP | 2.73 |
| 2 | pnma | $Ho_3Co$ | 0-D | AFM | 0 | WP, NL, DP | 3.42 |
| 3 | pnma | $Er_3Rh$ | 0-D | AFM | 0 | WP, NL, DP | 3.75 |
| 4 | pnma | $Ca_5Bi_3$ | 0-D | IM | 100 | WP, NL | 3.44 |
| 5 | pnma | $Ca_5Sb_3$ | 0-D | IM | 0 | WP, NL | 3.00 |
| 6 | pnma | $Sr_5Bi_3$ | 0-D | IM | 100 | WP, NL | 3.23 |
| 7 | pnma | $Sr_5Sb_3$ | 0-D | IM | 0 | WP, NL | 3.14 |
| 8 | P4/mbm | $Gd_2MgNi_2$ | 0-D | AFM | 0 | WP, NL, TDNP | 2.66 |
| 9 | P4/mbm | $K_2RbPt$ | 0-D | IM | 0 | WP, NL | 3.01 |
| 10 | P4/nmm | CeMnSi | 0-D | FM | 48 | WP, NL, DP | 4.47 |
| 11 | P4/nmm | CeMnGe | 0-D | FM | 35.2 | WP, NL, DP | 4.38 |
| 12 | $p4_2$/nmc | $Ba_2LiN$ | 0-D | IM | 7.3 | WP, NL, DP | 2.32 |
| 13 | I4/mmm | PrMgSn | 0-D | AFM | 19.1 | WP, TDNP | 4.01 |
| 14 | I4/mmm | NdMgSn | 0-D | AFM | 0 | WP, TDNP | 4.10 |
| 15 | I4/mmm | PrScGe | 0-D | AFM | 0 | WP, TDNP | 4.22 |
| 16 | I4/mmm | PrScSi | 0-D | AFM | 0 | WP, TDNP | 4.55 |
| 17 | I4/mmm | CeScGe | 0-D | AFM | 0 | WP, TDNP | 4.23 |
| 18 | I4/mmm | CeScSi | 0-D | AFM | 0 | WP, TDNP | 4.21 |
| 19 | I4/mmm | NdScSb | 0-D | FM | 82.7 | WP, TDNP | 3.90 |
| 20 | I4/mmm | TbScSb | 0-D | FM | 70.7 | WP, TDNP | 4.54 |
| 21 | I4/mmm | CeScSb | 0-D | FM | 48.2 | WP, TDNP | 3.95 |
| 22 | I4/mmm | NdScGe | 0-D | FM | 66.1 | WP, TDNP | 4.30 |
| 23 | I4/mmm | TbScGe | 0-D | FM | 22.2 | WP, TDNP | 4.02 |
| 24 | I4/mmm | NdScSi | 0-D | FM | 54.5 | WP, TDNP | 5.19 |
| 25 | $R\bar{3}m$ | LaCl | 0-D | FM | 67.1 | NL, AQHE | 3.99 |
| 26 | $R\bar{3}m$ | LaBr | 0-D | FM | 77.2 | NL, AQHE | 4.08 |
| 27 | $R\bar{3}m$ | YCl | 0-D | FM | 82.5 | NL, AQHE | 4.22 |
| 28 | $R\bar{3}m$ | YBr | 0-D | FM | 85.1 | NL, AQHE | 4.31 |
| 29 | $R\bar{3}m$ | ScCl | 0-D | IM | 80.2 | NL, AQHE | 4.30 |
| 30 | $R\bar{3}m$ | $Gd_2C$ | 2-D | FM | 23.1 | NL, AQHE | 3.33 |
| 31 | $P6_3$/m | $Ba_3MnN_3$ | 1-D | AFM | 0 | NL, AQHE | 2.45 |
| 32 | $P6_3$/m | $Sr_3MnN_3$ | 1-D | AFM | 0 | NL, AQHE | 2.59 |
| 33 | $P6_3$/mc | CaF | 0-D | IM | 77.0 | NL, NS, AQHE | 2.79 |
| 34 | $P6_3$/mc | SrF | 0-D | IM | 74.2 | NL, NS, AQHE | 2.62 |
| 35 | $P6_3/mc$ | BaF | 0-D | IM | 72.4 | NL, NS, AQHE | 2.56 |
| 36 | $P6_3/mc$ | CaBr | 0-D | IM | 68.0 | NL, NS, AQHE | 2.85 |
| 37 | $P6_3/mc$ | SrBr | 0-D | IM | 69.1 | NL, NS, AQHE | 2.75 |
| 38 | $P6_3/mc$ | BaBr | 0-D | IM | 70.2 | NL, NS, AQHE | 2.67 |
| 39 | $P6_3/mcm$ | $Ca_5Sb_3$ | 0-D | IM | 25.1 | WP, NS | 3.17 |
| 40 | $P6_3/mcm$ | $Ca_5Bi_3$ | 0-D | IM | 27.2 | WP, NS | 3.02 |
| 41 | $P6_3/mcm$ | $Sr_5Sb_3$ | 0-D | IM | 24.5 | WP, NS | 2.97 |
| 42 | $P6_3/mcm$ | $Sr_5Bi_3$ | 0-D | IM | 27.0 | WP, NS | 3.05 |
| 43 | $P6_3/mcm$ | $Ca_5As_3$ | 0-D | IM | 29.1 | WP, NS | 3.31 |
| 44 | $P6_3/mcm$ | $Sr_5As_3$ | 0-D | IM | 31.5 | WP, NS | 2.97 |
| 45 | $P6_3/mcm$ | $Ba_5As_3$ | 0-D | IM | 100 | WP, NS | 2.72 |
| 46 | $P6_3/mcm$ | $Ba_5Sb_3$ | 0-D | IM | 100 | WP, NS | 2.63 |
| 47 | $P6_3/mcm$ | $Ba_5Bi_3$ | 0-D | IM | 82.1 | WP, NS | 2.55 |
| 48 | $P6_3/mcm$ | $Yb_5Sb_3$ | 0-D | IM | 31.2 | WP, NS | 3.12 |
| 49 | $P6_3/mmc$ | $Yb_5As_3$ | 0-D | IM | 40.1 | WP, NS | 3.27 |
| 50 | $P6_3/mmc$ | $LaCl_2$ | 0-D | IM | 0 | WP, NS | 4.01 |
| 51 | $P6_3/mmc$ | $Ba_4Al_5$ | 0-D | IM | 4.4 | WP, NS, DP | 2.1 |

**TOC Graphic:**

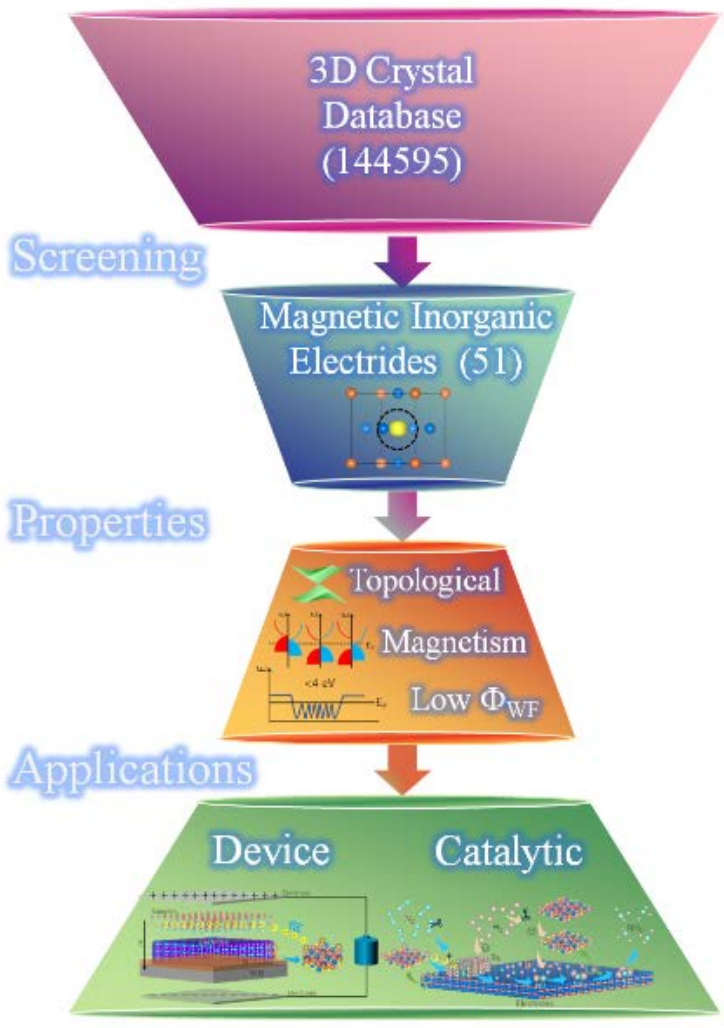